\documentclass[aps,twocolumn,floats,pre]{revtex4-1}
\usepackage{graphics,graphicx,epsfig,color}
\usepackage{amssymb,amsmath}
\usepackage{epsf,wrapfig}
\usepackage[mathcal]{euscript}

\usepackage{units} 
\usepackage{commath} 
\usepackage[]{titlesec}
\titleformat{\section}[block]{\color{black}\bfseries\filcenter}{}{1em}{}
\usepackage{hyperref}
\usepackage{upgreek} 

\usepackage[font=small,labelfont=bf,justification=raggedright, format=plain]{caption} 
\titlespacing\section{0pt}{10pt plus 2pt minus 0pt}{10pt plus 2pt minus 2pt}
\titlespacing\subsection{0pt}{5pt plus 2pt minus 0pt}{10pt plus 2pt minus 2pt}

\newcommand{\figtitle}[1]{\textbf{#1}}

\newcommand\figpanel[1]{(\textit{#1})~}
\newcommand\figpanelref[1]{\textit{#1}}

\renewcommand{\vec}[1]{\boldsymbol{#1}}

\newcommand{\ferr}{\ensuremath{f_{\mathrm{err}}}}
\newcommand{\W}{\ensuremath{\mathbf{W}}}
\newcommand{\mx}{\ensuremath{_{\mathrm{max}}}}
\newcommand{\foutl}{\ensuremath{f_{\mathrm{outline}}}}
\newcommand{\fpix}{\ensuremath{f_{\mathrm{pixel}}}}

\begin{document}

\title{Resolving coiled shapes reveals new reorientation behaviors in {\it C.~elegans}}


\author{Onno D. Broekmans\textsuperscript{a},
Jarlath B. Rodgers\textsuperscript{b,c},
William S. Ryu\textsuperscript{b,c,d},
Greg J. Stephens\textsuperscript{a,e} 
} 
\email{Corresponding author: g.j.stephens@vu.nl}
\affiliation{
\textsuperscript{a}Department of Physics and Astronomy, VU University Amsterdam, 1081 HV Amsterdam, The Netherlands\\ \textsuperscript{b}Donnelly Centre, University of Toronto, Toronto, ON M5S3E1, Canada\\ \textsuperscript{c}Department of Cell and Systems Biology, University of Toronto, Toronto, ON M5S 3G5, Canada\\\textsuperscript{d}Department of Physics, University of Toronto, Toronto, ON M5S1A7, Canada\\\textsuperscript{e}OIST Graduate Univeristy, Onna, Okinawa 904-0495, Japan
}


\begin{abstract}
We exploit the reduced space of {\em C. elegans} postures to develop a novel tracking algorithm which captures both simple shapes and also self-occluding coils, an important, yet unexplored, component of worm behavior. We apply 
our algorithm to  show that visually complex, coiled sequences are a superposition of two simpler patterns: the body wave dynamics and a head-curvature pulse.  We demonstrate the precise coiled dynamics  of an escape response and 
uncover new behaviors in spontaneous, large amplitude coils; deep reorientations occur through classical $\Omega$-shaped postures and also through larger, new postural excitations which we label here as $\delta$-turns.
We find that omega and delta turns occur independently, the serpentine analog of a random left-right step, suggesting a distinct triggering mechanism.  We also show that omega and delta turns display approximately equal rates and adapt to food-free conditions on a similar timescale, a simple strategy to avoid navigational bias.  
\end{abstract}
                                                                                                                                                                                                                                                                                                                                  
\keywords{}
\maketitle

\section*{Introduction} 
Much of our fascination with the living world, from molecular motors to the dynamics of entire societies, is with emergence ---  where the whole is surprisingly different than the sum of its parts (see, e.g.,~\cite{Laughlin:2014is}).
Yet, the existence of such collective organization also suggests that living systems, despite their enormous potential complexity, often inhabit only a much smaller region of their potential `phase space' and evidence for this lower-dimensional behavior is ubiquitous. For example, the motor control system produces movements that are far less complex than what the musculoskeletal system allows \cite{DAvella2003} and this hints at the presence of an organizational principle.  In a typical daily movement like walking, the central nervous system is thought to produce the full walking gait by combining low-level `locomotory modules', some of which appear to be universal among species \cite{Dominici2011}. Similarly, the dynamics in brain networks are organized in low-dimensional activity patterns \cite{Tkacik:2014em,Gao2015} and these patterns ---not individual neurons --- might be the carriers of information and computation \cite{Hopfield1982,Yoon2013}.  

The emergent dynamics of behavior, how animals move and interact, is particularly important as the ultimate function of the system \cite{Tinbergen:2005hk} and the scale on which evolution naturally applies.
Yet, our quantitative understanding of behavior is substantially less advanced than the microscopic processes from which it is produced, even as recent efforts have expanded this frontier \cite{Mirat:2013fx,Berman:2014ef,Cavagna:2014hd}.  How do we  analyze high-resolution behavioral dynamics and what does this reveal about an animal's movement strategy? How do we build effective models on the behavioral level where a `bottom-up' approach is daunting?  How do we connect analysis on the organism-scale to the properties of molecules, cells and circuits?  We approach these questions through the postural movements of the nematode {\em C. elegans}.

In {\em C. elegans}, the 2D space of body postures can be captured precisely and is also low-dimensional \cite{Stephens2008} so that the worm's motor behavior is fully quantified using a time series of only four variables.   These `eigenworms' are collective coordinates in the space of natural worm shapes and provide a notable reduction in complexity.  However, an important limitation of previous work is the inability to deduce the geometry of self-occluding body shapes.  In navigation, self-occluding body postures occur during  `omega turns' (a maneuver during which the worm's body briefly resembles the Greek letter $\Omega$) and are a general part of the worm's behavioral repertoire, ranging from foraging \cite{Stephens2010,Salvador2014}, and chemotaxis \cite{Pierce-Shimomura1999}, to escape from noxious stimuli \cite{Mohammadi2013}. For example, during escape behaviors worms use coiled shapes to reorient precisely by $180^\circ$ and the benefit seems obvious: it steers the worm back to safety. But how does a `blind' organism achieve this result without any visual reference to the outside world? While some of the neural and molecular mechanisms driving  omega turns have been uncovered \cite{Gray2005,Donnelly2013} and there has been previous work on crossed shapes \cite{Huang2006,Wang2009,Roussel2014,Nagy2015}, a quantitative analysis of such self-occluded posture dynamics is lacking.

Here, we exploit low-dimensionality to develop a novel and conceptually simple posture tracking algorithm able to unravel the worm's self-occluding body shapes. We apply our approach to analyze coiled shapes during two important behavioral conditions: the escape response induced by a brief heat shock to the head and spontaneous turns while foraging on a featureless agar plate.  We find that, in general, complex deep turn sequences can be viewed as a simpler superposition of body wave phase dynamics with a bimodal head swing followed by a unimodal curvature pulse.  In the escape response we show that while turning accounts for much of the $\sim180^\circ$ reorientation, the full distribution of reorientation angles is shaped by significant contributions from the reversal, turn and post-turn behaviors, a result consistent with the presence and action of the monoamine tyramine during the entire response. In natural crawling, the peak amplitudes of the curvature pulse reveal two distinct coiling behaviors --- the classical omega turn accomplishing large ventral-side reorientations and a previously uncharacterized `delta' turn which produces dorsal reorientations by overturning through the ventral side.  The omega and delta turns occur independently in time suggesting a separate triggering process but have similar rates, producing little overall bias in the trajectories.
\section*{Tracking posture with low-dimensional worm shapes} 

Previously, we analyzed movies of \textit{C.~elegans} freely crawling on an agar plate (Fig.~\ref{fig:InvertingTheTrackingProblem}A) \cite{Stephens2008}. For each movie frame, we identified the body of the worm, and applied a thinning algorithm to find the centerline. The worm's 2D body posture was characterized as a 100-dimensional vector of tangent angles along this centerline. (Fig.~\ref{fig:InvertingTheTrackingProblem}B--C). Principal Component Analysis revealed that more than 95\% of the variance in naturally-occurring body postures was captured by just four eigenvectors of the posture covariance matrix (Fig.~\ref{fig:InvertingTheTrackingProblem}D). As a result, any worm posture can be decomposed as a linear combination of these `eigenworms'  (Fig.~\ref{fig:InvertingTheTrackingProblem}E).  Worm behavior then becomes a smooth, low-dimensional trajectory through posture space (Fig.~\ref{fig:InvertingTheTrackingProblem}F). As an example, forward and backward crawling appear as approximately circular trajectories in the $(a_1, a_2)$ plane, and correspond to  limit-cycle attractors \cite{Stephens2008}.  However, for coiled shapes such as shown in Fig.~\ref{fig:InvertingTheTrackingProblem}H, the thinning algorithm does not produce a faithful reconstruction of the worm's actual posture (Fig.~\ref{fig:InvertingTheTrackingProblem}G). 

The above procedure can also be implemented in reverse to \emph{generate} worm images. For any point $\vec{p}$ in posture space (Fig.~\ref{fig:InvertingTheTrackingProblem}F), we can reconstruct the shape of the backbone (\figpanelref{G}). Knowing the thickness of the worm at each point along the body (which we estimate by averaging over many worm images), we are then able to draw a reconstructed body image (\figpanelref{H}; see Methods).  We then track the posture by finding, for each movie frame, the point in posture space (and thus the correct centerline) for which the reconstituted worm image is the most similar to the original image. This approach works for all worm postures --- in contrast to image thinning, which fails for self-overlapping shapes (Fig.~\ref{fig:InvertingTheTrackingProblem}H, inset).

Our `inverse' tracking algorithm consists of three basic elements.  (i) An image error function $\ferr$ quantifies how well a reconstituted worm image $\tilde{\W}(\vec{p})$ matches the movie frame $\W$ (Fig.~\ref{fig:TrackingWormPostures}A); (ii) an efficient optimization scheme to search for a global error minimum over all postures and; (iii) a method to resolve ambiguity as different self-occluding body shapes can give rise to the same image.  We measure image 
similarity using two specific shape metrics \cite{Yang2008}: outline shape, and coarse-grained pixel density (Fig.~\ref{fig:TrackingWormPostures}B).
By mapping this error function onto posture space: $\ferr(\vec{p}) = \ferr\left(\W, \tilde{\W}(\vec{p})\right)$, we create a fitness landscape, in which the position of the global minimum corresponds to the tracking solution 
and we find this minimum using a pattern search algorithm (a form of direct search \cite{Kolda2003}).  We retain multiple minima for each frame until a final step which minimizes total sequence error and we
sketch this process for a single mode in Fig.~\ref{fig:TrackingWormPostures}C.

\section*{Tracking reproduces both simple and self-occluding worm shapes with small errors}
Tracking results for a typical movie that includes complex, self-occluding shapes, are shown in Fig.~\ref{fig:TrackingWormPostures}D (see also Supporting Movie 1 and 2). In the gray rows at the top are the original movie frames; the reconstituted images from our inverse algorithm are below. While some minor inaccuracies are visible by eye, the overall result is remarkably similar.  To quantify posture tracking accuracy, we first compared the results of our algorithm to image thinning which allows for verification based on a large dataset. We used image thinning to construct a 100-dimensional vector of tangent angles $\vec{\theta}$, defined the tracking error as $\delta\theta = \left\Vert \vec{\theta}_{\mathrm{inv}} - \vec{\theta}_{\mathrm{thinning}} \right\Vert$, and we plot the distribution of these errors in Fig.~\ref{fig:TrackingWormPostures}E (black).  We also show the error in $\vec{\theta}$ that results from dimensionality reduction to the postural eigenmodes (black). Additionally, we show euclidean distances between tangent angle vectors of consecutive frames in a 16 Hz movie, representing limited time resolution (gray). For this dataset of non-crossed frames, our algorithm provides excellent performance, with tracking errors bounded by time resolution and dimensionality reduction. Even for errors in the tail of the distribution ($\delta\theta=\unit[3]{rad}$), backbones from the thinning and the `inverse' algorithm are quite similar (inset, gray backbones).  

A more relevant quantity for low-dimensional trajectories is the mode error $\delta a_i$ which is negligible for simple shapes as shown in Fig.~\ref{fig:TrackingWormPostures}F (yellow).  Finally, we created a dataset of self-overlapping body shapes for which backbones were manually drawn.  In  Fig.~\ref{fig:TrackingWormPostures}F (blue) we show that for the majority of crossed frames, the mode error is less than 10\% of the total range of naturally occurring mode values. As a visual reference, the reconstituted worm shapes corresponding to mode errors of $\delta a_i=1$ are shown in gray: these are noticeably flat.

\section*{Coiled dynamics in the escape response reveal precise reorientations and the superposition of the body wave and a head-curvature pulse}

We first applied our postural tracking algorithm to quantify the full shape dynamics of the {\it C. elegans } `escape response'. This is a stereotyped behavioral sequence, consisting of a pause, a reversal and an $\Omega$-turn, that quickly moves the worm away from a threatening stimulus.  Featuring only relatively simple coiled shapes, the escape response provided a useful test of our algorithm.  While recent work has connected the escape response with genetic, molecular, and neural mechanisms \cite{Donnelly2013}, the behavior itself has been described only qualitatively.  Here, we elicited an escape response by using an infrared laser pulse administered to the head of the worm, which raised the temperature by $\sim \unit[0.5]{^\circ C}$. 10 seconds of pre-stimulus behavior and 20 seconds of post-stimulus behavior were recorded at $\unit[20]{Hz}$. Each worm was only assayed once, to prevent adaptation. In total, $N=92$ worms were recorded, of which $N=91$ successful trackings were used in the final analysis.

A schematic of the response is shown in Fig.~\ref{fig:EscapeResponse}A, with the associated postural mode dynamics in Figs.~\ref{fig:EscapeResponse}(B,C). 
During normal, forward locomotion (i in Fig.~\ref{fig:EscapeResponse}A , $t <\unit[10]{s}$ in Fig.~\ref{fig:EscapeResponse}C), the worm crawls by propagating a sine-like wave through its body. This is reflected as a pair of phase-locked sinusoidal oscillations in $a_1$ and $a_2$ and we define the body wave phase angle $\varphi = -\arctan(a_2/a_1)$, where the minus sign ensures that $\dif \varphi / \dif t$ is positive during forward crawling. When the worm is stimulated by the infrared pulse (ii in Fig.~\ref{fig:EscapeResponse}A, pink line in Fig.~\ref{fig:EscapeResponse}C at $t=\unit[10]{s}$), it immediately backs up (iii), seen as a decrease in $\varphi$.  The end of this reversal and the beginning of the $\Omega$-turn is marked by a head-swing, visible as a bimodal pulse in $a_4$.  The $\Omega$-turn itself (iv) occurs as a large, unimodal pulse in $a_3$, and propagates head-to-tail.  This implies another switch of the direction of the body wave, and hence a return to increasing $\varphi$.  Finally, as the turn is finished, the worm resumes forward crawling (v).  The mode dynamics outlined above illustrate that the complexity of the escape sequence can be seen as a superposition of two simpler patterns: the body wave phase dynamics in $(a_1, a_2)$, and the head-curvature dynamics of $(a_3, a_4)$. A movie of these mode dynamics is available as Supporting Movie 3.

A notable feature of the escape response is how closely the worm controls its reorientation and our tracking algorithm also makes it possible to track the overall orientation continuously, across the different phases of the escape response.
In Fig.~\ref{fig:EscapeResponse}(D-E), we calculate how much each of the three response segments reorients the worm.  The distribution of reorientations for the full escape response is largely similar to the distribution during the omega-turn but includes contributions from the reversal and post-turn segments.  In the trial-averaged reorientation Fig.~\ref{fig:EscapeResponse}(E), we find  $\langle \Delta\theta\rangle = -0.89\pi \pm 0.05\pi \,\unit{rad}$ for the full response.  The omega turn itself results in $\langle \Delta\theta\rangle= -0.90\pi \pm 0.04\pi \, \unit{rad}$ while pre- and post-omega phases show smaller but significant contributions, $\langle \Delta\theta\rangle =0.13\pi \pm 0.03\pi \, \unit{rad}$ and $\langle \Delta\theta\rangle= -0.12\pi \pm 0.03\pi \, \unit{rad}$, respectively (errors are calculated using bootstrap across trials and are equivalent to standard errors in the mean).  Remarkably, the mean reorientation in the reversal and post-turn segments precisely cancel, suggesting a correction mechanism at the level of the average response so that the mean overall reorientation is entirely determined by the omega-turn.   No such precision is apparent in the variance where we find  $\delta \theta^2=0.69 \pi \pm 0.16 \pi \, \unit{rad}^2$ for the full response compared to the smaller  $\delta \theta^2=0.45 \pi \pm 0.16 \pi \, \unit{rad}^2$ for the turn segment.   Thus, while the omega turn is an effective maneuver for turning away from the stimulus, the full response orientation change is broadened by the reversal  ($\delta \theta^2=0.23 \pi \pm 0.05 \pi \, \unit{rad}^2$ ) and post-omega ($\delta \theta^2=0.19 \pi \pm 0.04 \pi \, \unit{rad}^2$) behaviors.

These observations allow us to hypothesize a subtle link between the behavior of the worm and the escape response at the neurotransmitter level  \cite{Donnelly2013}. As the worm enters the reversal phase, release of tyramine sets up an asymmetry in the worm's body and this appears as a baseline shift in the fluctuations of the third mode (see also Supporting Figure 1) leading to a positive bias in the reorientation, Fig.~\ref{fig:EscapeResponse}(D,E)\,(reversal). After the turn, lingering effects of the tyramine produce a similar baseline shift but as the worm is moving forward instead of backward, this now leads to an opposite orientation bias, Fig.~\ref{fig:EscapeResponse}(D,E)\,(post-omega).  

\section*{Coiled dynamics during foraging reveal distinct, large-amplitude turns}
To analyze more complex coiled shapes we applied our posture algorithm to foraging worm behavior on a flat agar plate (see Methods).  Under these conditions, worms navigate using a combination of maneuvers, including short and long reversals, pirouettes and even more continuous turns.  We are particularly interested in the pirouettes as they involve deep coils.  Such body bends are primarily encoded in the third postural eigenmode ($a_3$) and, as discussed in the previous section, peaks in $a_3$ are a characteristic feature of omega turns, and have a known role in reorientation of the worm \cite{Stephens2010}.

In Fig.~\ref{fig:SpontaneousTurns}A, we show the full distribution of postural mode  $a_3$ for all local extrema and note that the modes have been normalized so that negative $a_3$ amplitudes correspond to dorsal turns; ventral turns have strictly positive amplitudes. A clear asymmetry can be observed so that on top of a symmetric background distribution of shallow turns in both directions, we see, on the ventral side, two distinct additional peaks. Drawing reconstituted worm images for the center values of these two peaks, it is clear that the peak at $a_3 \sim 15$ corresponds to a `classic' $\Omega$ shape. The second peak, at $a_3 \sim 23$, shows a body shape with a much higher characteristic curvature. In Fig.~\ref{fig:SpontaneousTurns}A\,(right), we have `folded' the dorsal side of the distribution over the ventral side, highlighting the ventral asymmetry at high $a_3$ amplitudes. As noted in the figure, we refer to turns in the lower-amplitude peak as \emph{omega turns} and distinguish these from the higher-amplitude \emph{delta ($\delta$) turns} in the second peak. As for the omega turn, the name \emph{delta turn} is chosen to reflect the $\delta$-like shape of the worm during a typical sequence.

Returning to the original tracking movies, the presence of these two classes of turns is clearly visible. In Fig.~\ref{fig:SpontaneousTurns}B we display movie stills for two example turns: one omega turn, and one delta turn. During the classical omega turn, the worm slides its head along its body, similar to the escape response, ending up with large, primarily ventral reorientation. A delta turn, on the other hand, is much deeper: the worm completely crosses its head over its body, resulting in an dorsal reorientation by `over-turning' across the ventral side.

\section*{Delta and omega turns are complementary components of a navigational strategy}

In postural dynamics, the delta and omega turns differ primarily in their $a_3$ pulse amplitude; their turn kinematics are otherwise very similar (Supporting Figure 2).  However, when turns do occur, they result in a dramatically different change of overall orientation.  As in the escape response we use our posture algorithm to track the worm's overall body reorientation and in Fig.~\ref{fig:SpontaneousTurns}C, we show how the worm reorients using both omega (orange) and delta (blue) turns. Simply put, omega turns reorient the worm by large, ventral angles, while delta turns reorient the worm dorsally by `over-turning'  through the ventral side. The difference in reorientation angle may provide a hint as to why these two behaviors  exist. Earlier, we saw that the neural mechanisms that produce the escape-response omega turn are fundamentally asymmetric, producing only ventral turns (through disinhibition of the VD motor neurons) \cite{Donnelly2013}. If the worm uses the same neural infrastructure during free crawling, this would only ever allow it to reorient itself towards its ventral side. Lacking a dorsal `copy' of the same neural infrastructure, the worm could instead hyper-activate the existing infrastructure to produce ventral `over-turning'. These `over-turns' are what we call delta turns, and enable the worm to also reorient towards its dorsal side. We also find that delta and omega turns occur seemingly independently; the mutual information between time-binned, time-shifted series for both turning event time series has a maximum of less than a few percent (see Methods and Supporting Figures 3-4).  On the other hand, evidence that the turns can be jointly controlled is shown in Fig.~\ref{fig:SpontaneousTurns}D. Here, we plot the frequency of turning events over the course of the experiment (see Methods). As the worm searches for food in a larger area, the turn frequency decreases significantly --- a well-known phenomenon \cite{Gray2005,Maricq2005,Srivastava2009}- and both omega and delta turns show similar frequencies and adaptation.

\section*{Discussion}

The ability to track self-overlapping shapes of {\it C. elegans} together with the eigenworm projection of postures, provides a complete and quantitive accounting of the worm's locomotory behavior in 2D.   Among living systems with a nervous system, such an exact behavioral description is  unique and is likely to be especially important as new techniques emerge for the simultaneous imaging of a substantial fraction of the worm's neurons during free behavior \cite{Nguyen:2015dp,Venkatachalam:2015bx}.  Our posture tracking algorithm itself is conceptually simple and relies on an optimized image search within the low-dimensional space of worm shapes.  Indeed, while the identification of low-dimensionality occupies an important role in quantitative approaches to living systems (see e.g. \cite{Machta:2013ga,Daniels:2015ht,Ganguli:2012bx}), here we have leveraged these dimensions to elucidate important and previously unknown aspects of {\em C. elegans} coils.  Interestingly, we were able to apply the characterization of body postures found previously for non--self-overlapping body shapes \cite{Stephens2008}, to capture shapes that \emph{do} self-overlap; even the simpler eigenworm space allows for substantial posture diversity.     

We applied our tracking algorithm to two important behaviors, an evoked escape response and the deep, spontaneous turns that occur during foraging.  Viewing the coiled turn as a trajectory through the low-dimensional posture space, a simple model emerges: a superposition of the body wave (a circular trajectory in posture space corresponding to simple forward and backward crawling), and coupled pulses along the third and forth mode (corresponding to the deep coil and a preceding head oscillation). This model is consistent with the molecular mechanisms found to orchestrate the escape response \cite{Donnelly2013}.   Our results also hint at a possible answer to how precise reorientations of $180^\circ$ are accomplished: the worm could use its own body as a `guide' for reorientation. During the omega-turn,  the distribution of $a_3$ peak amplitudes (Fig.~\ref{fig:EscapeResponse}D\,(Omega turn, inset)) lies close to a value of 15: the lowest $a_3$ value that generates a self-touching body shape. This suggests that the worm might have evolved to coil until it just intersects its own body, which it then slides along to find its way back.

While the omega turn has previously been considered as a single class of \textit{C.~elegans} behavior, our analysis of the amplitudes of the curvature mode $a_3$ pulses associated with deep coils reveals the presence of two distinct subpopulations. `Classic' omega turns, featuring the signature $\Omega$ body shape, reorient the worm to the ventral side while delta turns reorient the worm dorsally by over-turning through the ventral side.  We show that these deep dorsal and ventral reorientations occur independently in time with approximately equal rates, which is important if there is to be no overall bias in the foraging trajectories.  

While distinct in visual appearance, omega and delta turns differ only in the amplitude of the curvature mode and we have shown that these behaviors are discretely separable during foraging.  However, coiling is also observed in other contexts, including a variety of mutants \cite{Yemini:2013bd,Nagy2015}.  We expect our methods will be useful in analyzing such shapes and as  a guide for uncovering neural and molecular mechanisms underpinning coiling behavior.

Deep turns and reorientations form an important component of the taxis strategy of {\em C. elegans} \cite{Pierce-Shimomura1999,Gray2005, Stephens2010,Salvador2014}.  Under foraging and chemotaxis conditions, these behaviors are seemingly stochastic \cite{Srivastava2009,Gallagher2013}, producing a broad distribution of reorientation angles analogous to tumbling in  the bacteria {\em E. coli} \cite{Berg:1972wt}.  However, unlike bacterial tumbling (which occurs through an instantaneous switch in the rotation direction of a molecular motor and the resulting unbundling of the flagellar tail, see e.g \cite{Berg:2006bn}) the worm's reorientation is driven by a long, controlled sequence of stereotyped postural changes.  Thus an important question is how does the worm effectively randomize its direction?  We have shown here that half the variability in {\em C. elegans} foraging reorientations is due simply to the initial random choice of delta or omega turns.  However, even the level of stochasticity can be modulated, as evidenced by the largely deterministic reorientation in the escape response, differing response variability depending on the strength of a thermal stimulus \cite{Mohammadi2013} and the slow adaptation of the reversal rate \cite{Gray2005,Stephens2011}.  Overall, such a combination of behaviors, flexible and stochastic combined with patterned and deterministic is likely to observed even in more complex organisms, including humans. In initiating the detailed analysis of {\em C. elegans} turning behavior we hope that our work here offers first step towards a general understanding of these processes.

\section*{Methods}
{\noindent {\bf Data:} We used two datasets encompassing both foraging and escape response behavioral conditions. The foraging data was explored previously \cite{Stephens2011} and for more details on data collection see also \cite{Stephens2008}. In short, \textit{C.~elegans} N2-strain worms were imaged with a video tracking microscope at $f=\unit[32]{Hz}$. Worms were grown at $\unit[20]{^\circ C}$ under standard conditions \cite{Sulston1974}. Before imaging, worms were removed from bacteria-strewn agar plates using a platinum worm pick, and rinsed from \textit{E.~coli} by letting them swim for 1~minute in NGM buffer. They were then transferred to an assay plate (9-cm Petri dish) that contained a copper ring (5.1-cm inner diameter) pressed into the agar surface, preventing the worm from reaching the side of the plate. Recording started approximately 5 min after the transfer, and lasted for $\unit[2\,100]{s}$ (35~min). In total, data from $N=12$ worms was recorded.  The second dataset, the `escape response' condition, was recorded following procedures as described in ref.~\cite{Mohammadi2013}. In short, worm recordings took place in a temperature-controlled room ($\unit[22.5]{^\circ C} \pm \unit[1]{^\circ C}$). A 100-ms, 75-mA infrared laser pulse from a diode laser ($\lambda=\unit[1\,440]{nm}$) was administered to the head of the worm, raising the temperature in a FWHM-radius of $\unit[220]{\upmu m}$ by $\sim \unit[0.5]{^\circ C}$. 10 seconds of pre-stimulus behavior and 20 seconds of post-stimulus behavior were recorded at a frame rate of $\unit[20]{Hz}$. Each worm was only assayed once, to prevent adaptation. In total, $N=92$ worms were recorded, of which $N=91$ successful trackings were used in the final analysis.
\bigskip

{\noindent \bf{Image processing and shape reconstruction:}} All movie frames were converted to binary mages and cropped, using standard image processing functions in MATLAB (R2014b, The Mathworks, Natick, MA) \cite{Stephens2008}. For faster processing, before analysis with the inverse tracking algorithm, the foraging data was down-sampled to $\unit[16]{Hz}$ by dropping every second frame. To reconstitute an image of a worm with a body posture $\vec{p}=(a_1, \ldots, a_5)$, we first calculated the vector of backbone tangent angles from $\vec{\uptheta} = \sum_{i} p_i \vec{\hat{e}}_i$, with $\vec{\hat{e}}_i$ the $i$'th eigenworm. Knowing the total arc length $l$ of the worm, we could calculate the position of each of the 100 points along the backbone. At each backbone point $j$, we then drew a filled circle with radius $r_j$ to capture the worm's body thickness (see also Fig.~\ref{fig:InvertingTheTrackingProblem}g,h) and thus create the worm image. Circle radii $r_j$ for a particular worm were computed from movies of uncrossed worm postures for that specific worm. In each such frame, after finding the centerline (backbone) and outline of the worm \cite{Stephens2008}, we could find $r_j$ as the minimum distance between backbone point and outline. This was averaged across all frames. Similarly, the total arc length $l$ of the worm was computed by averaging across frames. For the error function described below, the overall orientation of the worm in the image is important and we generate images of worms in all possible orientations by adding an overall orientation value $\langle\theta\rangle\in[0,2\pi)$ to the backbone tangent angle vector. This gives us a full backbone vector $\vec{\uptheta}_{\mathrm{F}} = \langle\theta\rangle + \sum_{i=1}^{5} a_i \vec{\hat{e_i}}$.

\bigskip
{\noindent \bf{Image error function and inverse algorithm:}} The shape error function compares two binary worm images $\W_1$ and $\W_2$ and is computed as $\ferr = \foutl \cdot \fpix$.  For $\foutl$ we calculate a set of tangent angles $\psi$ to the perimeter of the worm shape (Fig.~\ref{fig:TrackingWormPostures}b, left). We find the 4-connected outline of the worm in the binary image $\W_i$, fit a spline through these points, and discretize it into 201 segments sampled at equal arc length. The 200 resulting angles between the segments form a vector $\vec{\psi}_i = (\psi_{i,1}, \psi_{i,2}, \ldots, \psi_{i,200})$; the total length of the segments is $\ell_i$. $\foutl$ is now $\foutl = C_0  \lvert \vec{\psi}_1 - \vec{\psi}_2 \rvert^2 + C_1 \left( \ell_1 - \ell_2 \right)^2$, for arbitrary constants $C_0$ and $C_1$. Note that the value of $\foutl$ is sensitive to the choice of starting points for tracing the 4-connected outline in each image; this is resolved by choosing the pair of starting points that minimizes $\foutl$. For $\fpix$, we first align the images $\W_1$ and $\W_2$ so that their centroids overlap. Each image is then divided into a grid of 10x10-pixel `blocks' (Fig.~\ref{fig:TrackingWormPostures}b, right). For each block $(j,k)$ ($j=1, \ldots, n$; $k=1, \ldots, m$) in image $\W_i$, the fraction $d_i(j,k)$ of black pixels in the block is calculated. This coarse-graining into blocks allows for, e.g., minor inaccuracies in the generation of worm images from mode values, without affecting the error function. We then calculate $\fpix$ as $\fpix = \frac{1}{nm} \sum_{j,k} \left( d_1(j,k) - d_2(j,k) \right)^2$. In earlier trials, we found that using five postural eigenmodes gave us significantly better tracking results than only using four. Since our error function is sensitive to the overall rotation of the worm, we amended the five-dimensional posture space with an extra dimension for the overall orientation $\langle\theta\rangle$. This means that the search space for our algorithm is six-dimensional, with 5 postural dimensions, and 1 rotational dimension.  To find a tracking solution for a frame, we ran 580 pattern searches (using MATLAB's patternsearch function) from randomly distributed starting points in search space, with the error function described above as objective function. Only solutions with an error value less than $1.0$, a threshold value obtained through trial-and-error, were kept. Solutions within a given hypercube of dimensions $ [3.0, 3.0, 3.0, 3.0,  2.5]$ were merged, leaving only the solution with the lowest error value. This finally resulted in zero, one, or more potential tracking solutions per movie frame. To speed up the optimization, we applied two additional constraints. Firstly, we bounded the absolute value of the eigenmodes to $(18,18,34,12,6)$, for each of the five modes respectively. We verified that the distributions of eigenvalues $a_i$ found in our tracking data tailed off before reaching these limits. Secondly, we set a limit to the maximum local curvature of the worm's backbone so that elements in the resulting theta vector that are 10 indices apart must not be different by more than 1.95 rad.
This limit rules out body shapes that were unnaturally coiled.

Importantly, we note that  our inverse problem is fundamentally ill-posed: multiple body postures may produce the same two-dimensional worm image (e.g.~Fig.~\ref{fig:TrackingWormPostures}B, bottom) and for each movie frame $j=1,\ldots,N$, we generally find multiple potential  solutions which we label $\{\vec{p}_j^k\}$, with $k=1,\ldots,M_j$. Even for simple, non-crossed postures, there can be two solutions ($M_j=2$) corresponding to the swapped locations of the head and tail.  Across the movie, we label the indices of the correct solutions as a vector $\vec{b} = (b_1, \ldots, b_N)$ .  We explicitly allow $b_j=0$ in case the optimization process fails and use a cubic spline to interpolate across any such gaps. Let us call the point in posture space for movie frame $j$, resulting from this interpolation step, $\tilde{\vec{p}}_j (\vec{b})$. To find  $\vec{b}^*$ for the full, correct tracking solution of the movie we seek the solution vector, $\vec{b}^*$ that minimizes the total sequence error $E(\vec{b}) = \sum_{j=1}^{N} \ferr \left[ \W_j, \tilde{\W}\left( \tilde{\vec{p}}_j (\vec{b}) \right) \right]$ and we constrain the distance between two successive frames to be below $\vec{v}\mx$, a continuity constraint which simply reflects the fact that the worm can only change its posture by a maximum amount between two movie frames.  

\bigskip
{\noindent {\bf{Tracking Pipeline:}} In a first pass of the data, the `classic' worm tracking algorithm based on image skeletonization was used on all frames \cite{Stephens2008}. This fast algorithm yields high-accuracy tracking results for frames with simple, non--self-overlapping body shapes. It also automatically labeled crossed frames.  For the foraging dataset, the data was cut into smaller segments to allow for faster parallel processing. Each segment consisted of a series of non-crossed frames, followed by a series of crossed frames, followed by more non-crossed frames. This effectively segmented the data by omega turn (936 segments in total for the 12 worm trajectories). For the escape response dataset, such segmentation was not necessary, due to the smaller size of the data for each worm. Frames that were previously labeled as `crossed' were tracked using the inverse algorithm described above. The result was an interpolated, smooth trajectory through posture space. When using this pipeline as-is, the algorithm would occasionally swap the locations of head and tail between frames. To resolve head/tail orientation correctly throughout a segment, we implemented four steps. (1)~During the filtering and interpolation step, we allowed the algorithm to pick, for each non-crossed frame, not just the solution given by the `classic' algorithm; it could also pick an alternative version in which head and tail were swapped (this version can be trivially computed). (2)~We explicitly included a limit for the maximum change of overall orientation $\langle\theta\rangle$ between frames of $\sim \pi$ rad per second in the maximum velocity vector $\vec{v}_{\mathrm{max}}$. Any head/tail swaps between frames violate such a maximum change of $\langle\theta\rangle$. (3)~After the filtering and interpolation step had produced a full tracking solution, we computed the error for both that tracking solution, as well as a version in which the head and tail were swapped for all frames in the segment. This fixed the overall head/tail orientation for the full segment. (4)~As a final check, we manually verified and, if necessary, corrected head/tail orientations during post-processing.

\bigskip
{\noindent \bf{Tracking quality:}} In total, $92$  escape responses and $936$ free-crawling segments (each containing one self-overlapping turn; see Methods) were analyzed. The escape response tracking results were inspected manually, and $91$ trackings ($99\%$) were considered successful, as they were visually close to the appearance of the original worm. For the free crawling dataset,instead, after inspection of a representative sample of $236$ segments across multiple worms, $96\%$ were estimated to be successful.
First, we assessed the quality of our tracking algorithm for non-crossed worm shapes (Fig.~\ref{fig:TrackingWormPostures}D). We used both the `classic' algorithm and the `inverse' algorithm to track $N=15433$ non-crossed frames from the foraging dataset. For each frame, we calculated the euclidean distance between the two resulting $\vec{\theta}$ vectors giving the `inv. tracking' distribution in Fig.~\ref{fig:TrackingWormPostures}a (black). In the same figure, the `dim. reduct.' distribution (blue) was calculated from euclidean distances between the full $\vec{\theta}$ vector from the classic algorithm, and $\vec{\theta}_{\mathrm{reduct}} = \sum\limits_{i=1}^{5} a_i \vec{\hat{e}}_i$, where $\vec{\hat{e}}_i$ are the eigenworms.  This represents the information lost in only using the first five postural eigenmodes (which capture ${>}95\%$ of the shape variance \cite{Stephens2008}). The `time res.' distribution (yellow) represents the euclidean distance between $\vec{\theta}$ vectors from consecutive frames in a movie. In Fig.~\ref{fig:TrackingWormPostures}(E), we additionally collected a dataset of four movies, featuring visually distinct types of omega turns. For the $N=348$ crossed frames in these four movies, backbones were hand-drawn on the worm images, independently from the tracking results. We considered these manual tracking results a `gold standard', and compared them to the final results of our inverse tracking / filtering and interpolation algorithms. The resulting mode errors $\delta a_i$ are plotted as the blue/dark distributions. We also incude the mode errors for the set of $15433$ non-crossed frames (yellow).
\bigskip

{\noindent \bf{Definition of large-amplitude turns:}} For the escape response data, the largest peak in $a_3$ between $t=\unit[10]{s}$ (the time of the stimulus) and $t=\unit[29]{s}$ was identified as the apex of the omega turn. To locate the end of the omega turn, the first zero of $a_4$ after the apex was found; any point after that root that had $a_3<3$ was considered to be the end of the omega turn. This ensured that the negative peak in $a_4$, representing a high-curvature state of the tail at the end of the omega turn, had finished, and that the worm had reached a relatively `straight' shape. For such straight shapes, the overall orientation $\langle\theta\rangle$ has a straightforward, intuitive interpretation. The same criterion was used, in the opposite direction, to find the start of the omega turn. If no starting point and/or end point of the omega turn could be found, the recording was excluded from the analysis. (In the escape response dataset, this was the case for 15 out of 91 recordings). We used the same criterion to find both omega and delta turns in the foraging condition. For detection of local extrema in $a_3$, a standard peak-finding algorithm was used to detect both minima and maxima (based on the MATLAB findpeaks function, which defines a peak as a data point with a greater value than its immediate neighbors). Only extrema with a minimum prominence of $0.5$ were kept. Some $a_3$ peaks featured smaller sub-peaks in their shoulders; such sub-peaks were discarded.

\bigskip 
{\noindent \bf{Orientation:}} Orientation changes were computed by comparing the overall orientation $\langle\theta\rangle$ between two reference points around each omega or delta turn. The apex of each deep turn was the largest $a_3$ peak identified previously. The first reference point was the last frame before the turn's apex that featured a `straight' body shape --- i.e., a body shape with a low maximum local curvature. Only for such relatively `flat' worm shapes does the overall orientation $\langle\theta\rangle$ correspond directly to the intuitive orientation assigned to the worm. Similarly, the second reference point was the first frame after the turn's apex with such a straight body shape.  Importantly, our postural tracking algorithm allows us to continuously follow the orientation angle through coiled shapes and this is important for identifying the 'overturning' reorientation effects of delta turns. For the analysis of the worm's reorientation during the escape response (Fig.~\ref{fig:EscapeResponse}\,D,E), $N=91$ escape responses were analyzed. Each 30-second recording was segmented by first finding the omega turn. After identification of the omega turn, the reversal phase was simply defined as the first frame after the stimulus with a negative body wave phase velocity $\dif \varphi / \dif t$, up until the start of the omega turn. The `post-omega' phase was any data after the end of the omega turn until the end of the recording at $t=\unit[30]{s}$.   If any of the chosen recordings did not have a successfully detected omega turn, it was skipped. 

\bigskip
{\noindent \bf {Mutual information between omega and delta-turn event time series:}} To calculate the mutual information between the omega and delta turns during foraging, we created binary event time series by first identifying the time of the $a_3$ peak and then binning these times into bins of width $2$, $4$, $10$, or $\unit[20]{s}$.  We then calculated the mutual information between these binary time series as in ref.~\cite{Strong:299311}. The mutual information was calculated for different relative shifts, ranging from $-60$ to $\unit[+60]{s}$. Mutual information across time shifts never exceeded $\sim 3\%$ of the maximum entropy of each time series indicating that these turns occur independently. See also Supporting Figure 3.

\bigskip
{\noindent \bf{Omega and delta turn frequency adaptation:}} In Fig.~\ref{fig:SpontaneousTurns}D, we show how the average turn frequencies for omega and delta turns change over the course of the 35-minute foraging experiments. Turns were detected by using the peak detection algorithm outlined above. Using the amplitude boundaries identified in Fig.~\ref{fig:SpontaneousTurns}b, $a_3$ extrema with an absolute value between 10 and 20 were classified as `omega turns', while extrema with an absolute value greater than 20 were considered to be `delta turns'. We also distinguished between ventral turns, with a positive amplitude, and dorsal turns, with a negative amplitude. We counted the average number of turns per unit time, across the 12 experiments, in a 10-minute sliding window, shifted across the data in 5-minute steps. The first 200 seconds of each experiment were discarded. The total of these extrema consists of three-populations: a tail of the symmetric distribution of `shallow turns', and two types of ventral deep turns, the delta and omega turns. To find the number of omega turns we therefore counted the number of $a_3$ peaks with an amplitude between $-20$ and $-10$ in each time window, and subtracted this from the total number of $a_3$ peaks with an amplitude between $+10$ and $+20$. An identical procedure with $|a_3|>20$ gives the number of delta turns which is in excellent agreement with the number of omega turns, differing only to the extent expected from Poisson number fluctuations.

\acknowledgments{We thank GJ Berman, AEX Brown and TS Shimizu for discussions and SURFsara (www.surfsara.nl) for help with the Lisa Compute Cluster.  ODB was supported by start-up funds from the Department of Physics and Astronomy, Vrije Universiteit Amsterdam.  GJS acknowledges funding from the Department of Physics and Astronomy, Vrije Universiteit and The Okinawa Institute of Science and Technology Graduate University.   WSR and JBR thank The National Science and Engineering Council of Canada (NSERC).}
\bibliography{refs}
\bibliographystyle{pnas2011}

\begin{figure*}
\includegraphics[width=1.6\columnwidth]{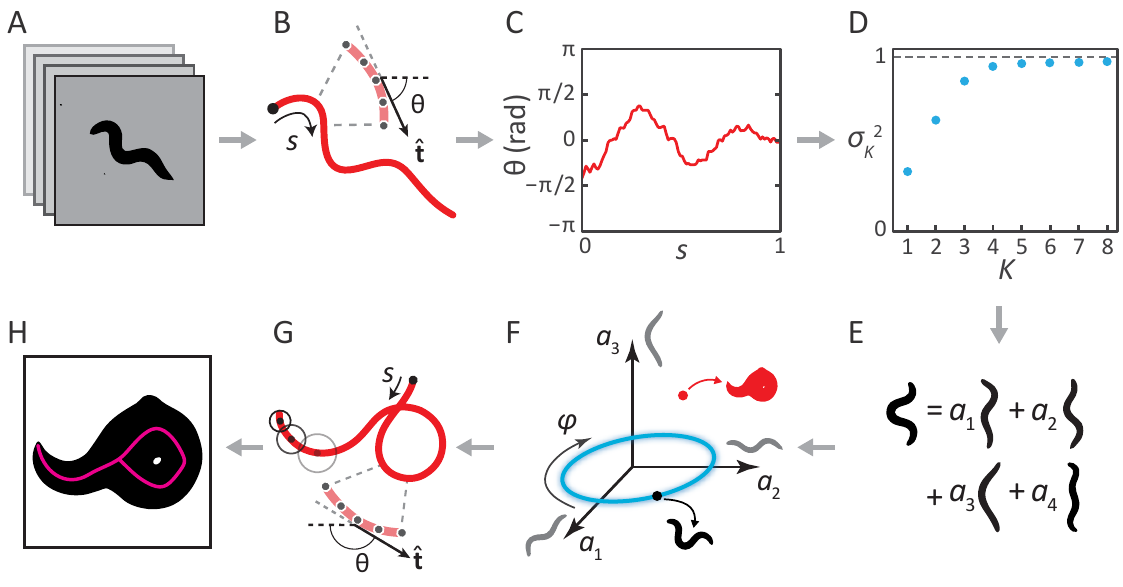}
\caption{\figtitle{Inverting posture analysis to generate worm images.} (A-E) We previously showed that the space of \textit{C.~elegans} body postures is low-dimensional. \figpanel{A}For a set of images of a freely moving worm, \figpanel{B}we find the centerline of the body using image thinning (black point indicates the head). \figpanel{C}At equidistant points along the centerline, we measure the direction $\theta(s)$ of the tangent $\hat{t}$. After subtracting $\langle\theta\rangle$, this gives a description of the worm's shape that is intrinsic to the worm itself. \figpanel{D}Principal Component Analysis reveals that only four eigenvectors of the shape covariance matrix are needed to account for $\sim 95\%$ of the variance in $\theta(s)$. \figpanel{E}Hence, any body shape can be decomposed as a linear combination of postural `eigenworms'. \figpanel{F}Alternatively, we can think of any body posture as a point in a low-dimensional `posture space', spanned by the eigenworms (gray). Forward crawling is then represented by clockwise progression along a circular trajectory in the $(a_1,a_2)$ plane (blue oval, body wave phase angle $\varphi$). \figpanel{G}For any point in this space, we can easily calculate the shape of the backbone. \figpanel{H}A series of filled circles with radii representing the worm's thickness, are used to draw an image of the worm's body, inverting the original postural analysis to generate an image. For self-overlapping shapes such as  in \figpanelref{H}, image thinning (\figpanelref{H, magenta)} does not produce an accurate reconstruction of the posture (\figpanelref{G}, red).}
\label{fig:InvertingTheTrackingProblem}
\end{figure*}

\begin{figure*}
\centerline{\includegraphics[width=1.6\columnwidth]{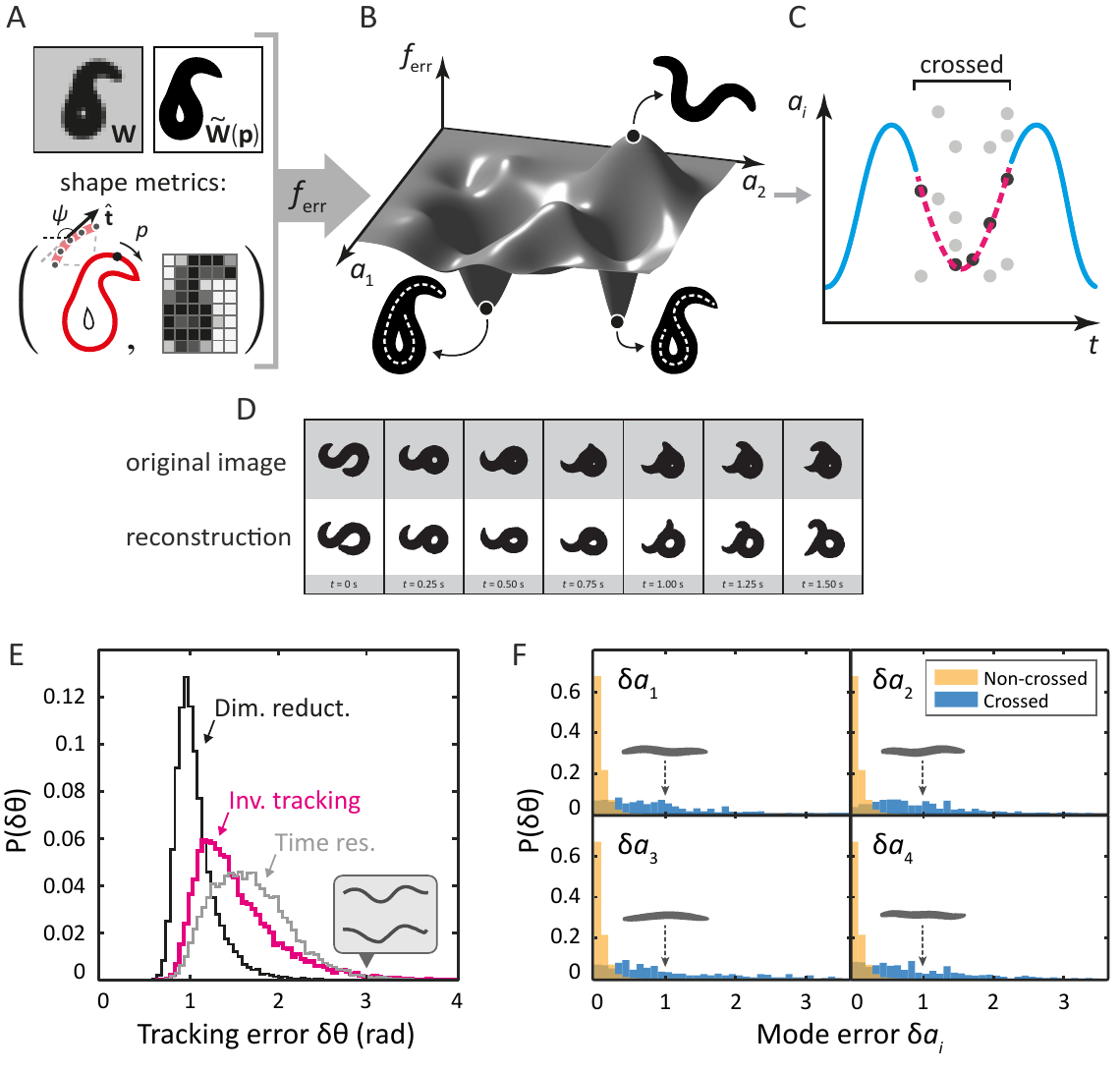}}
\caption{\figtitle{Tracking coiled shapes by searching for image matches in posture space.} {\bf Top: tracking algorithm sequence.} \figpanel{A}For each movie frame $\mathbf{W}$ and reconstituted worm image $\mathbf{\tilde{W}}(\vec{p})$ for posture $\vec{p}$ we apply two metrics, one based on the shape of the boundary (left), and one based on a coarse-grained pixel density matrix (right).
\figpanel{B}An error function $\ferr$ based on these two shape metrics generates a fitness landscape (schematically shown). The position of the global minimum of $\ferr$ corresponds to the tracking solution; if a frame is ambiguous, multiple minima may be present. \figpanel{C} For non-crossed body postures, a simple image thinning algorithm suffices to obtain time series of the modes $a_i$ (blue line, schematically shown). For crossed frames, we use the procedure outlined in \figpanelref{A--B}. Due to the inherent ambiguity of such images, multiple solutions are generally found for each frame (light gray points). Using the filtering algorithm described in the main text, we identify the correct solutions (dark gray points). The resulting smooth trajectory (magenta, dotted line) forms the full tracking solution. \figpanel{D} Sample tracking results (bottom, while background), contrasted with original images (top, gray background) for a turning sequence. {\bf Bottom: the inverse algorithm accurate tracks both simple and coiled worm shapes with small error.} \figpanel{E}Histogram of tracking errors for non--self-overlapping worm shapes, quantified as the euclidean distance $\delta\theta$ between the tangent angle vector $\vec{\theta}$ from our algorithm, and $\vec{\theta}$ found by image thinning (magenta). For scale, the error due to dimensionality reduction to the postural eigenmodes is shown in black. We also show the euclidean distance between $\vec{\theta}$ in consecutive frames, representing the confidence in $\vec{\theta}$ due to the finite time resolution of the movie (gray). Even for an extreme value of $\delta\theta=\unit[3]{rad}$ (gray arrow), backbones from the `classic' algorithm (top) and our algorithm (bottom) are nearly indistinguishable by eye (inset).
\figpanel{F}Tracking error in eigenmode values for the first four modes. For uncrossed worm shapes (yellow/light), our algorithm shows negligible tracking errors.  For a smaller set of crossed frames, we compare to a manually found solution (blue/dark). For scale, we show reconstituted images for worms with a single nonzero mode value of $a_i=1$; these `error worms' are essentially flat.}
\label{fig:TrackingWormPostures}
\end{figure*}

\begin{figure*}
\centerline{\includegraphics[width=1.5\columnwidth]{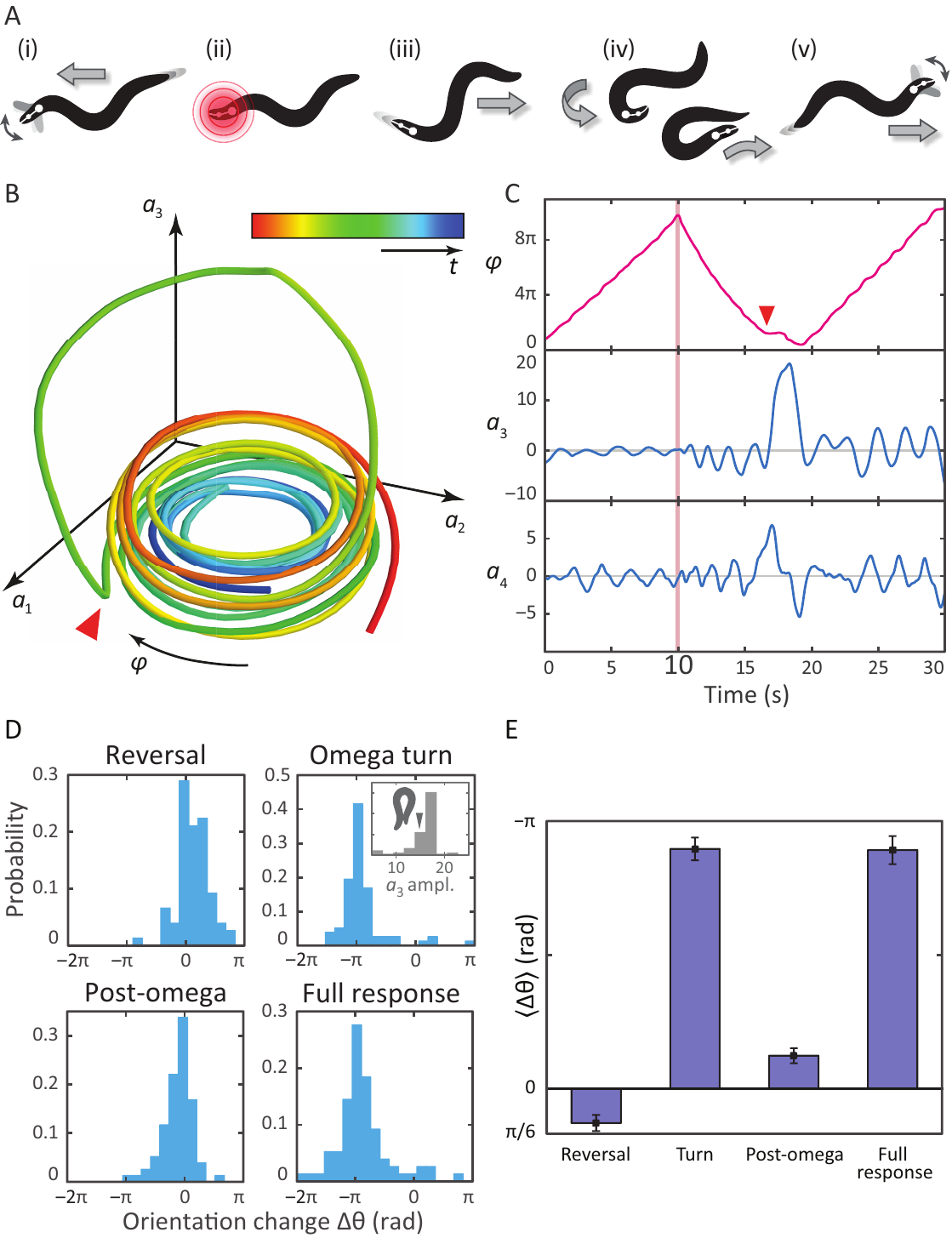}}
\caption{\figtitle{Tracking coiled postures and reorientation in the escape response}.
\figpanel{A}Schematic overview \cite{Donnelly2013} with worm body shapes extracted from tracking data: \textit{i}~forward locomotion and exploratory head motions; \textit{ii}~infrared laser stimulus; \textit{iii}~reversal phase; \textit{iv}~omega turn; \textit{v}~resumption of forward locomotion in the opposite direction.
\figpanel{B}Trajectory through posture space; $\varphi$ indicates direction of increasing body wave phase angle and color encodes time with blue for $t=0$ and red at $t=\unit[30]{s}$. The reorientation coil is evident as a large excursion along the third mode, starting at the red arrow.
\figpanel{C}The same trajectory as in \figpanelref{B}, in terms of the body wave phase angle $\varphi$ and the postural modes $(a_3,a_4)$. The heat shock occurs at $t=\unit[10]{s}$ (pink bar).  The omega turn is initiated by a head swing as seen in $a_4$ followed by a large pulse in $a_3$ and is linked to a `re-reversal', a return to forward movement. \figpanel{D} An important feature of the escape response is the change in the worm's overall orientation and we apply our algorithm to track this reorientation for each response segment.  While turning  accounts for much of the reorientation, the full response distribution is shaped by significant contributions from all three segments. In particular, the small but biased reorientations of the reversal and post-turn segments originate in the $a_3$ fluctuations outside of the turn (see  the time series in \figpanelref{C} and also Supporting Figure 1) and are consistent with the release and presence of the monoamine tyramine during the entire response.  \figpanel{E}  The precision of the escape response is evident in the trial-mean reorientation $\langle \Delta \theta \rangle$. The mean reorientation in the reversal and post-turn segments closely cancel, suggesting a correction mechanism at the level of the average response.  In the inset to \figpanel{D, Omega turn} we also show the distribution of $a_3$ amplitudes and this is peaked near coiled shapes in which the worm barely touches-perhaps indicative of a strategy in which the worm uses it's own body to achieve $\sim 180^\circ$  reorientations.}
\label{fig:EscapeResponse}
\end{figure*}

\begin{figure*}
\centerline{\includegraphics[width=1.6\columnwidth]{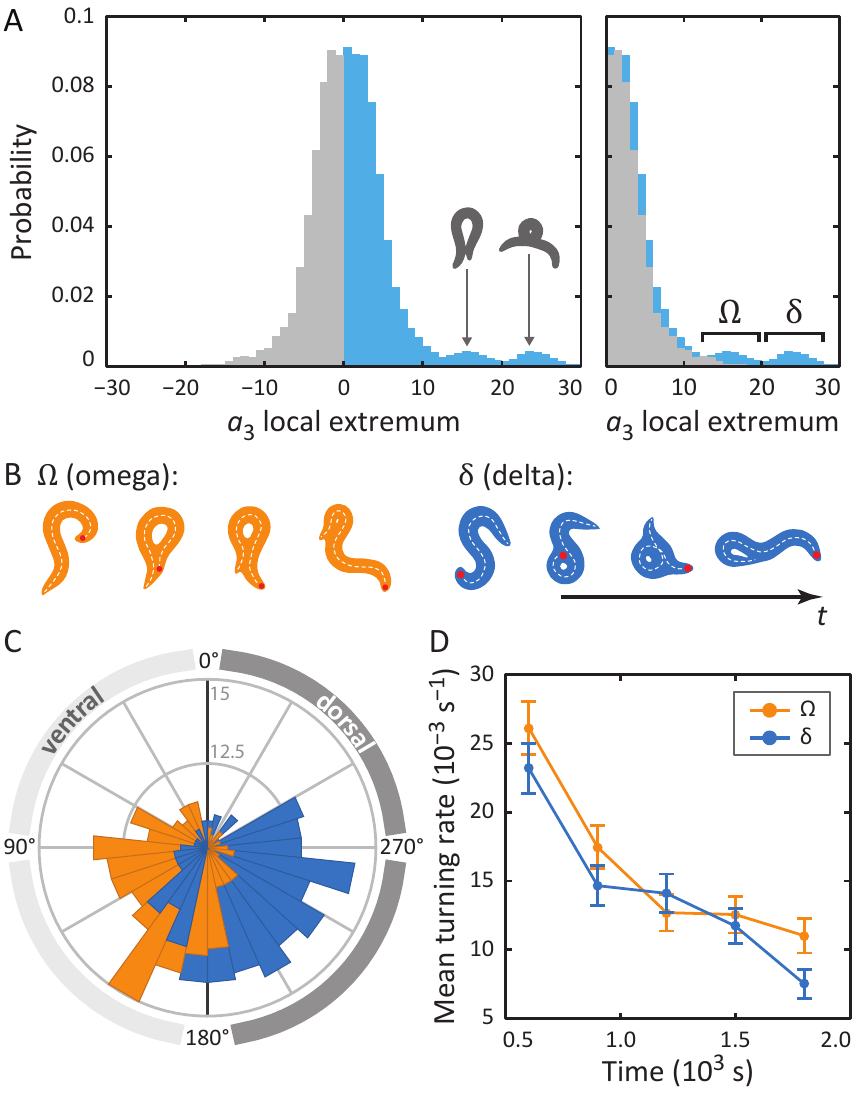}}
\caption{\figtitle{Unraveling coiled shapes during foraging reveals two distinct ventrally-biased classes of large-amplitude turns}. \figpanel{A}(left) Probability of the amplitude of all local extrema in the time series of the third postural eigenmode $a_3$. Colors represent the sign of the $a_3$ amplitude, and hence the dorsal (gray) or ventral (blue) direction of the resulting turn. 
\figpanel{A}(right) As previously, with all negative $a_3$ amplitudes now plotted as positive. The peaked excess in the distribution for large ventral bends correspond to `classic' $\Omega$ (omega) shapes, and previously undescribed deeper $\delta$ (delta) turns. Insets in \figpanelref{A} (left) show reconstructed worm shapes for the indicated $a_3$ amplitudes. In \figpanelref{B} we show stills from a movie of a worm making a classical omega turn (left, yellow), and a deep `delta' turn (left, blue). The head is marked with a red dot and the dashed lines indicate postures determined from our inverse tracking algorithm .  The dynamics of $\delta$-turns are largly similar to $\Omega$-turns differing primarily in the the amplitude of the bending mode $a_3$ and the overall time to complete the maneuver (see Supporting Figure 2).
\figpanel{C}Histogram of orientation change ($\Delta\langle\theta\rangle$) due to ventral omega turns (yellow/light) and ventral delta turns (blue/dark). Ventral reorientations are accomplished through omega-turns. To reorient to the the dorsal side, however, {\em C. elegans} employs delta-turns which  `over-turn' through the ventral side. \figpanel{D }Average turning rate during the tracking experiment. Ventral omega and delta turns are temporally independent, suggesting a separate triggering mechanism, but occur with approximately equal rates that adapt similarly with time spent away from food, a simple strategy to avoid any dorsal-ventral navigational bias.}
\label{fig:SpontaneousTurns}
\end{figure*}


\begin{figure*}
\centering
\includegraphics[width=1.4\columnwidth]{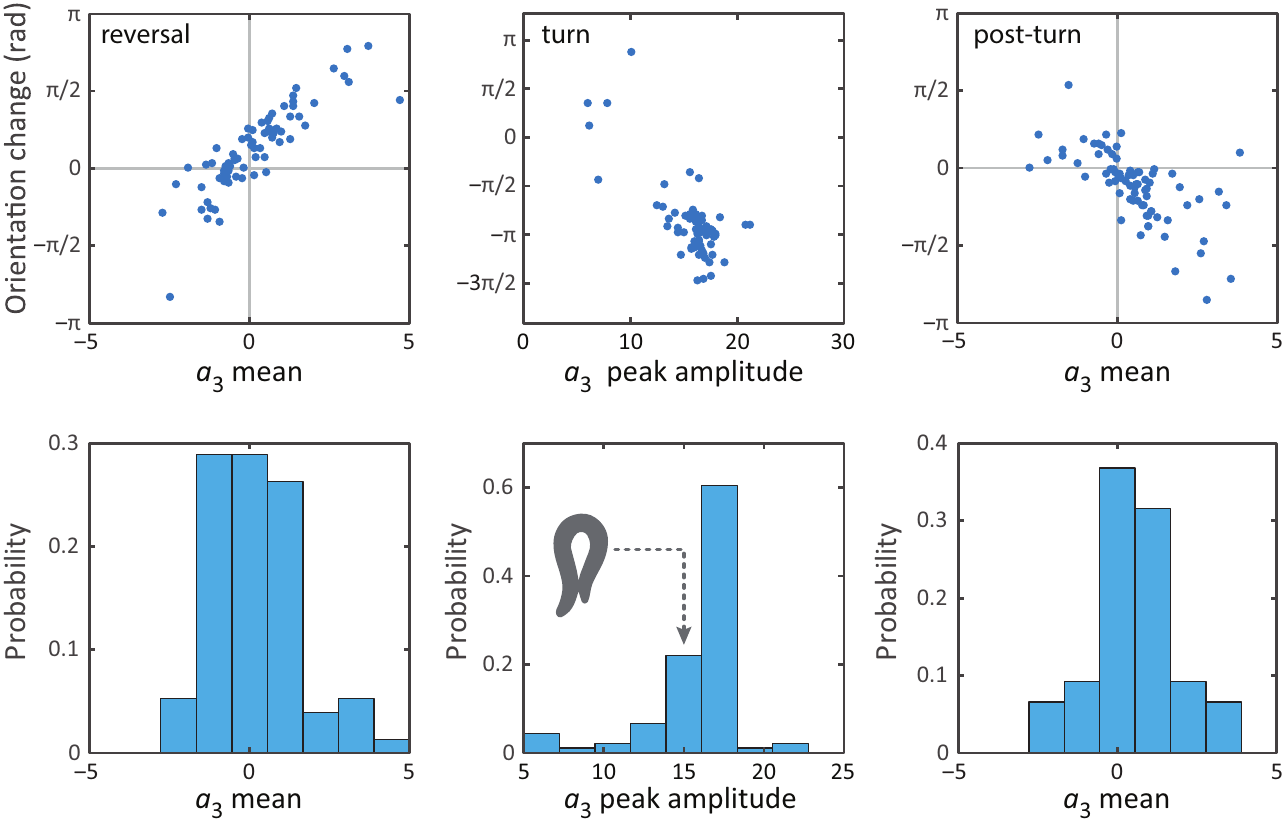}
\caption*{{\bf Supplementary Figure 1:} Bias in the turning mode $a_3$ and resulting reorientation occurs during all behaviors of the escape response. From previous work on the interpretation of the postural eigenmodes, we know that the third eigenmode (an overall bending of the worm) is linked to reorientation of the worm \cite{Stephens2008,Stephens2010}. We therefore tested if any asymmetry in the fluctuations of $a_3$ during the reversal phase could be linked to the observed reorientations. Such asymmetry is also visible in Fig.~3C as a baseline shift of the third mode during the reversal. (top) The mean $a_3$ value, versus the resulting orientation change, during the reversal and post-omega behaviors, respectively. The orientation change is strongly correlated with the mean $a_3$ value. (bottom) Peak amplitude of the $a_3$ peak corresponding to the omega turn, versus the resulting orientation change. Histogram of mean $a_3$ values, during the reversal phase and post-omega phase, respectively, showing similarly asymmetric distributions. Histogram of $a_3$ peak amplitudes during the omega turn. We also show a reconstituted worm image for an $a_3$ value of 15, for which the worm is barely self-occluded.
}
\end{figure*}

\begin{figure*}
\centering
\includegraphics[width=1.25\columnwidth]{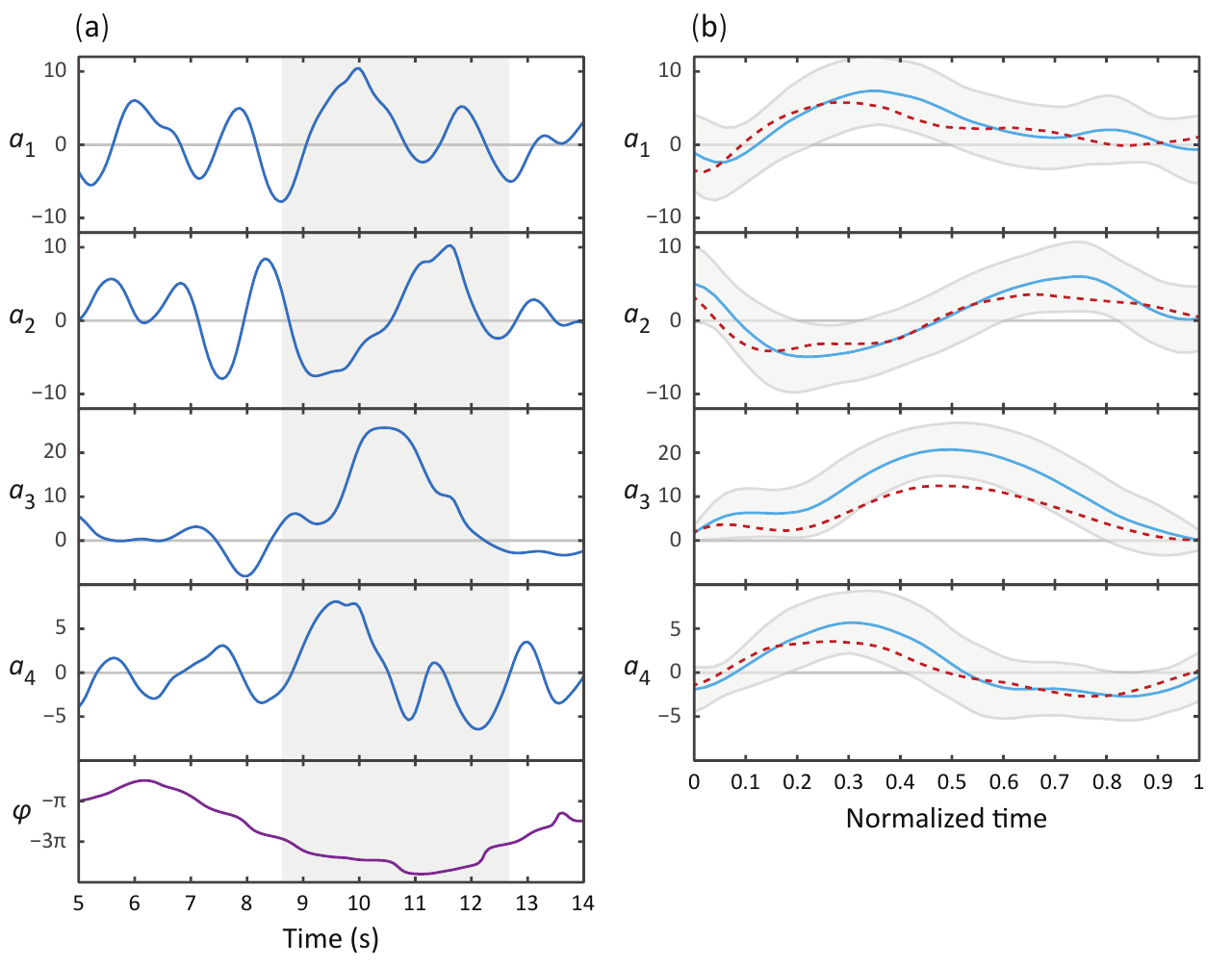}
\caption*{{\bf Supplementary Figure 2:} Omega and delta turns follow similar kinematics; while visually quite distinct, the primary difference is the amplitude of the curvature pulse $a_3$. (a) Typical time series for the postural eigenmodes $a_{1..4}$ during a deep delta turn; $\varphi$ is the body wave phase angle, as defined in Fig.~1F. Shaded area indicates the delta turn, as defined in the Methods. (b) Average eigenmode time series during a delta turn (blue, $N=348$). Gray lines indicate SD. For comparison, the average escape response omega turn is also shown (red dotted line). Time has been normalized with respect to the total length of the turn: $\unit[6]{s} \pm \unit[2]{s}$ (mean $\pm$ SD) for delta turns, $\unit[7]{s} \pm \unit[3]{s}$ for escape-response omega turns.
}
\end{figure*}

\begin{figure*}
\centering
\includegraphics[width=1.25\columnwidth]{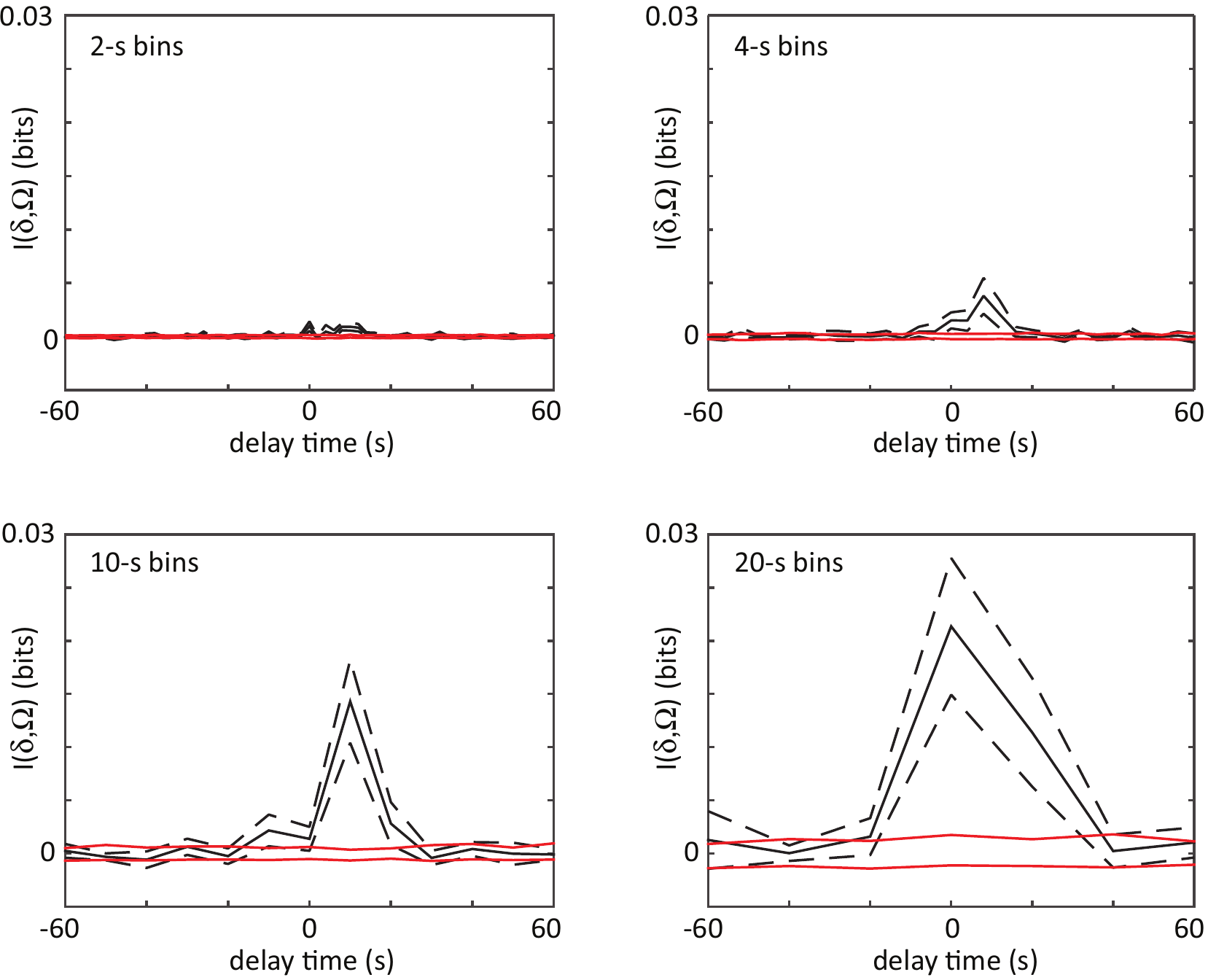}
\caption*{{\bf Supplementary Figure 3:} The shifted mutual information between $\delta$-turn and $\Omega$-turn time series.  Events are localized by the large amplitude peaks in the curvature mode and bins from 2 to 20 seconds are used to convert these timings into a binary event series.  The mutual information is very small in all cases (for scale the entropy of either event series is $\sim$ 1 bit) indicating that the two different turn types occur independently. Dashes are estimated errors from finite sampling and red lines denote the mutual information between shuffled time series which would be zero in the infinite data limit. }
\end{figure*}

\begin{figure*}
\includegraphics[width=1.25\columnwidth]{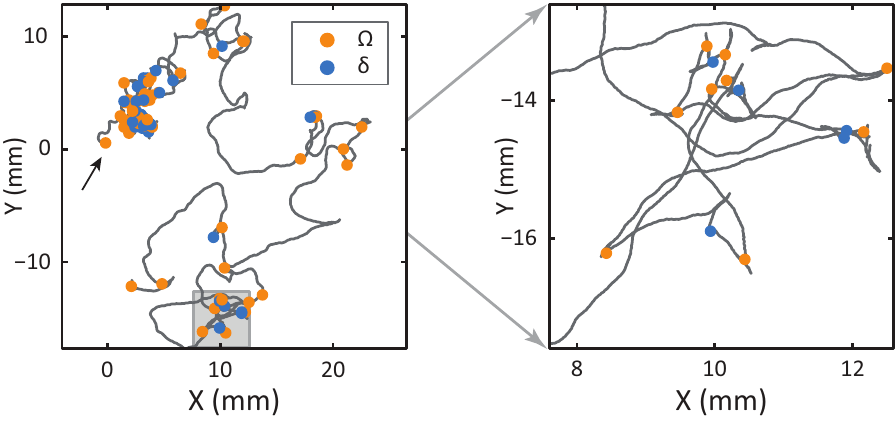}
\captionsetup{labelformat=empty}
\caption*{{\bf Supplementary Figure 4:} (left) Location of one of the 12 tracked worms over the course of a 35-minute tracking experiment (off-food), starting at $(0,0)$ (black arrow). $\Omega$-turns (orange) and $\delta$-turns (blue) are highlighted. (right) An enlargement of the area marked in gray.  While both turns occur more frequently at early times, there is no correlated pattern as consistent with their independence. }
\end{figure*}

\begin{figure*}
\captionsetup{labelformat=empty}
\caption*{{\bf Supplementary Movies:}\\
 (1) \href{https://dl.dropboxusercontent.com/u/26988873/LeveragingLowDimensions/EscapeResponse.avi}{\color{blue}https://dl.dropboxusercontent.com/u/26988873/LeveragingLowDimensions/EscapeResponse.avi\color{black}}. Tracking results for the escape response.  Left images are data while at the right are reconstructed images from our tracking algorithm. \\
 (2)  \href{https://dl.dropboxusercontent.com/u/26988873/LeveragingLowDimensions/SpontaneousComplex.avi}{\color{blue}https://dl.dropboxusercontent.com/u/26988873/LeveragingLowDimensions/SpontaneousComplex.avi\color{black}}. Tracking results for a complex, spontaneous coil.  Left images are data while at the right are reconstructed images from our tracking algorithm. \\
  (3) \href{https://dl.dropboxusercontent.com/u/26988873/LeveragingLowDimensions/EescapeResponseModes.mp4} {\color{blue} https://dl.dropboxusercontent.com/u/26988873/LeveragingLowDimensions/EescapeResponseModes.mp4\color{black}}. The dynamics of the escape response in the space of the first three eigenworms.  On the right we show the full body posture which turns red at the moment of the thermal impulse.  On the left are the dynamics in mode space.  The large-amplitude omega turn is visible as a `figure-8' trajectory. Note that even during the turn the body wave is advancing.  In general, turning behavior is a superposition of the body wave and curvature dynamics.
}
\end{figure*}

\end{document}